\documentclass[useAMS,usenatbib]{mn2e}
\usepackage{natbib}
\usepackage{color,graphicx} 
\usepackage{amsmath}        
\usepackage{amsfonts}       
\usepackage{amssymb}        

\title[Optical pulsations from SGR\,0501+4516]{The first observation
  of optical pulsations from a soft gamma repeater: SGR\,0501+4516}
\author[V. S. Dhillon et al.]{V. S. Dhillon,$^{1}$\thanks{E-mail: vik.dhillon@sheffield.ac.uk} 
T. R. Marsh,$^{2}$ S. P. Littlefair,$^{1}$ C. M. Copperwheat,$^{2}$ 
\newauthor 
R. D. G. Hickman,$^{2}$ P. Kerry,$^{1}$ A. J. Levan,$^{2}$
N. Rea,$^{4}$ C. D. J. Savoury,$^{1}$ 
\newauthor 
N. R. Tanvir,$^{3}$  R. Turolla,$^{5,6}$ K. Wiersema$^{3}$ \\
$^{1}$Department of Physics and Astronomy, University of Sheffield, Sheffield S3 7RH, UK \\
$^{2}$Department of Physics, University of Warwick, Coventry CV4 7AL, UK \\
$^{3}$Department of Physics and Astronomy, University of Leicester, Leicester LE1 7RH, UK \\
$^{4}$Institut de Ci\'encies de l'Espai (CSIC--IEEC), Campus UAB, Torre C5-parell, 2a planta, 08193, Barcelona, Spain \\
$^{5}$Dipartimento di Fisica, Universit\`a di Padova, via F. Marzolo 8, I-35131 Padova, Italy \\
$^{6}$Mullard Space Science Laboratory, University College London, Holmbury St. Mary, Dorking, Surrey, RH5 6NT, UK}

\begin{document}

\date{Submitted for publication as a Letter on 2011 May 13.}

\maketitle

\begin{abstract}
  We present high-speed optical photometry of the soft gamma repeater
  SGR\,0501+4516, obtained with ULTRACAM on two consecutive nights
  approximately 4 months after the source was discovered via its
  $\gamma$-ray bursts. We detect SGR\,0501+4516 at a magnitude of $i'
  = 24.4\pm0.1$. We present the first measurement of optical
  pulsations from an SGR, deriving a period of $5.7622\pm0.0003$\,s,
  in excellent agreement with the X-ray spin period of the neutron
  star. We compare the morphologies of the optical pulse profile with
  the X-ray and infrared pulse profiles; we find that the optical,
  infrared and harder X-rays share similar double-peaked morphologies,
  but the softer X-rays exhibit only a single-peaked morphology,
  indicative of a different origin. The optical pulsations appear to
  be in phase with the X-ray pulsations and exhibit a root-mean-square
  pulsed fraction of $52\pm7$\%, approximately a factor of two greater
  than in the X-rays. Our results find a natural explanation within
  the context of the magnetar model for SGRs.
\end{abstract}


\section{Introduction}

Soft gamma repeaters (SGRs) derive their name from the manner in which
they are usually discovered -- the emission of repeated `short bursts'
of hard X-rays/soft $\gamma$-rays with durations of 0.1$-$0.2\,s and
luminosities of 10$^{38}$$-$10$^{41}$\,erg\,s$^{-1}$.  Less
frequently, SGRs emit even more energetic
(10$^{41}$$-$10$^{43}$erg\,s$^{-1}$) bursts of longer duration
(1$-$60\,s), known as `intermediate flares', and rarer still, `giant
flares' (10$^{44}$$-$10$^{47}$erg\,s$^{-1}$, 300$-$600\,s), during
which, for a fleeting instant, the SGR outshines the entire
Galaxy. For a review of SGRs, see~\cite{woods06}~and~\cite{mereghetti08}.

As well as periods of bursting/flaring, SGRs also exhibit persistent
X-ray emission, from which it has been deduced that SGRs are all
isolated neutron stars with relatively long spin periods in the range
2$-$9\,s, relatively rapid spin-down rates of
$\sim$10$^{-10}$$-$10$^{-12}$\,s\,s$^{-1}$ and high inferred dipole
magnetic field strengths of $B$$\sim$10$^{14}$$-$10$^{15}$\,G
(however, see \citealt{rea10b}). The loss of rotational energy is
insufficient to explain the persistent X-ray luminosity and hence an
alternative energy source is required. The most-commonly accepted
model is the decay of the strong magnetic field. This so-called
`magnetar' model can explain both the bursts/flares and the persistent
emission of the SGRs, the former via cracking of the neutron-star
crust induced by the drifting field, and the latter via internal
heating and magnetospheric currents induced by the drifting/twisting
field (\citealt{thompson95}, \citealt{thompson02}).  Alternative
theories to the magnetar model do exist, the most prominent being the
`fall-back disc' model (\citealt{vanparadijs95},
\citealt{chatterjee00}, \citealt{alpar01}). In this model, the
persistent X-ray emission from SGRs, and the closely-related Anomalous
X-ray Pulsars (AXPs), is powered by the accretion of material left
over from the supernova explosion that formed the neutron star.

Only 21 SGRs and AXPs are currently
known\footnote{McGill~SGR/AXP~Online~Catalogue:\newline
  http://www.physics.mcgill.ca/$\sim$pulsar/magnetar/main.html}, 7 of
which are confirmed as SGRs (although the distinction between SGRs and
AXPs is becoming increasingly blurred, e.g. \citealt{rea09b}). The
discovery of each new source is therefore of considerable scientific
interest. On 2008 August 22, the {\em Swift} Burst Alert Telescope
(BAT), observed multiple, short ($<128$\,ms) bursts of soft
($<100$\,keV) $\gamma$-rays from a source close to the Galactic plane
\citep{barthelmy08}. The source's position, and the duration, energy,
and repetition of the bursts, led to the classification of the object
as a soft gamma repeater, SGR\,0501+4516 \citep{barthelmy08}; at the
time, this was the first new SGR to be discovered for a
decade. Subsequent observations of the X-ray afterglow allowed the
spin period \citep{gogus08} and rate of period decrease
\citep{woods08a} to be measured, further supporting the source's
classification as an SGR. Detailed studies of SGR\,0501+4516 in the
X-rays have been published by \cite{rea09b} and \cite{gogus10}. No
radio counterpart of the SGR was detected \citep{hessels08}, but a
faint, possibly variable, object coincident with the X-ray afterglow
was discovered in the infrared \citep{tanvir08} and optical
\citep{fatkhullin08}. In this paper, we present the results of
follow-up, time-resolved optical observations of SGR\,0501+4516,
obtained with the aim of understanding the emission processes at work.

\vspace*{-0.05cm}
\section{Observations and data reduction}
\label{sec:obsred}

\begin{figure}
 \centering
 \includegraphics[clip,width=5.75cm,angle=270]{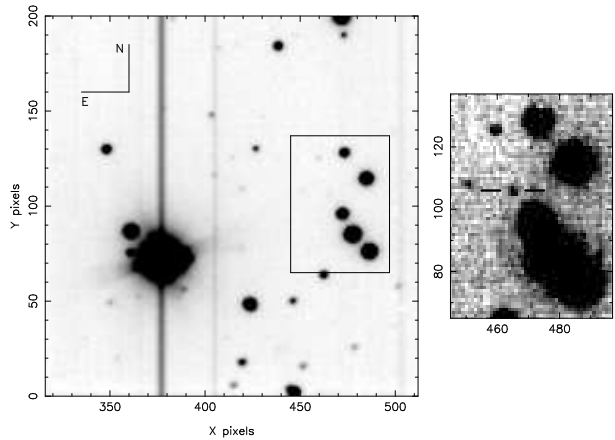}
 \caption{$i'$-band finding chart for SGR\,0501+4516. Left: Summed
   $i'$-band image of the field around SGR\,0501+4516 on 2009 January 1,
   with a total exposure time of 3247\,s. The box shows the portion of
   the field that is plotted at a higher contrast on the right. The
   orientation of the image is marked on the upper left-hand side. The
   pixel scale is 0.3\,$^{\prime\prime}$/pixel, hence the field of view in this image is
   60$^{\prime\prime}\times$60$^{\prime\prime}$. The vertical streaks are due to the frame transfer
   process. Right: Higher-contrast plot of a 16$^{\prime\prime}\times$22$^{\prime\prime}$ field around
   SGR\,0501+4516, highlighting the detection of the pulsar in the $i'$
   band. The X-ray error circle lies within the seeing disc of the
   pulsar and is hence invisible.}
  \label{fig:fc}
\end{figure}

We observed the proposed optical counterpart of SGR\,0501+4516 using
the high-speed, triple-beam camera ULTRACAM \citep{dhillon07b} on the
4.2\,m William Herschel Telescope on La Palma. We obtained 10\,152
$u'$, $g'$ and $i'$ frames, each of 0.475\,s exposure time (a run
length of 1.4\,hr), on 2008 December 31, and a further 6836 such
frames (a run length of 1.0\,hr) 26 hours later on 2009
January 1. ULTRACAM was used in two-windowed mode, giving a dead time
between each frame of 0.024\,s, where each frame is time-stamped to a
relative (i.e. frame-to-frame) accuracy of $\sim$50\,$\mu$s and to an
absolute accuracy of $\sim$1\,ms using a dedicated GPS system
\citep{dhillon07b}. Observations of the SDSS standard G\,243$-$38
\citep{smith02} were also obtained to flux calibrate the data. Both
nights were photometric with seeing of
0.8$^{\prime\prime}-$1.2$^{\prime\prime}$ and first-quarter Moon. The
sum of the $i'$-band frames obtained on 2009 January 1 is shown in
Figure~\ref{fig:fc}.

The data were reduced using the standard ULTRACAM data reduction
software \citep{dhillon07b}. All frames were first
debiased and then flat-fielded, the latter using the median of
twilight sky frames taken with the telescope spiralling. Adopting the
same successful approach that we used in our studies of the AXPs
4U\,0142+61 \citep{dhillon05} and 1E\,1048.1$-$5937 \citep{dhillon09},
we extracted light curves of SGR\,0501+4516 using two different
techniques:

\subsection{Technique (i)}
\label{sec:tech1}

First, we extracted a light curve of SGR\,0501+4516 by shifting each
frame to correct for image motion and then adding them into 10
evenly-spaced phase bins using the X-ray epoch, spin period and
spin-down rate of \cite{rea09b}. This ephemeris is accurate enough to
cover the epoch of our optical observations and is in good agreement
with the original X-ray ephemeris reported by \cite{israel08}. An
optimal photometry algorithm \citep{naylor98} was then used to obtain
sky-subtracted counts from SGR\,0501+4516 and an $i'$$\sim$13
comparison star $\sim$28$^{\prime\prime}$ to the east of the SGR, the
latter acting as the reference for the profile fits and
transparency-variation correction (although this correction made only
a negligible difference to the light curves). The position of
SGR\,0501+4516 relative to the comparison star was determined from a
sum of all the images, and this offset was then held fixed during the
reduction so as to avoid aperture centroiding problems. The sky level
was determined from a clipped mean of the counts in an annulus
surrounding the target stars and subtracted from the object counts.

\subsection{Technique (ii)}
\label{sec:tech2}

A light curve was also obtained using an independent method, where we
omitted the phase binning step described in Section~\ref{sec:tech1}
and simply performed optimal photometry on the 16\,988 individual data
frames, followed by a search for the X-ray period in a periodogram.

\section{Results}
\label{sec:results}

\subsection{Magnitudes}
\label{sec:mags}

Figure~\ref{fig:fc} shows that we have clearly detected an object in
the $i'$ band with magnitude $i'$\,=\,24.4$\pm$0.1 at position
RA$_{\rm J2000} = 05^{\rm h} 01^{\rm m} 06.73^{\rm s}, \delta_{\rm
  J2000} = 45^{\circ} 16^{\prime} 34.1^{\prime\prime} (\pm 0.3^{\prime\prime})$, in excellent agreement
with the {\em Chandra} X-ray position of SGR\,0501+4516
\citep{gogus10}. The source remained undetected in the $u'$ and $g'$
bands to 3$\sigma$ limiting magnitudes of $u' > 24.7$ and $g' > 26.9$,
as expected due to the high hydrogen column density ($N_{\rm
  H}\sim10^{22}$\,cm$^{-2}$, \citealt{rea09b}) at the low galactic
latitude of the SGR. The $i'$-band magnitude of the source in our
observations is fainter than the values of $I = 23.3\pm0.4$ on 2008
August 25 \citep{fatkhullin08} and $i' = 23.5\pm0.4$ on 2008 September
2 \citep{ofek08}, indicating that the SGR had declined in brightness
by approximately one magnitude in the 4 months since ouburst.

\subsection{Optical light curves}
\label{sec:lc_opt}

\begin{figure}
 \centering
 \includegraphics[clip,width=13.0cm,angle=270]{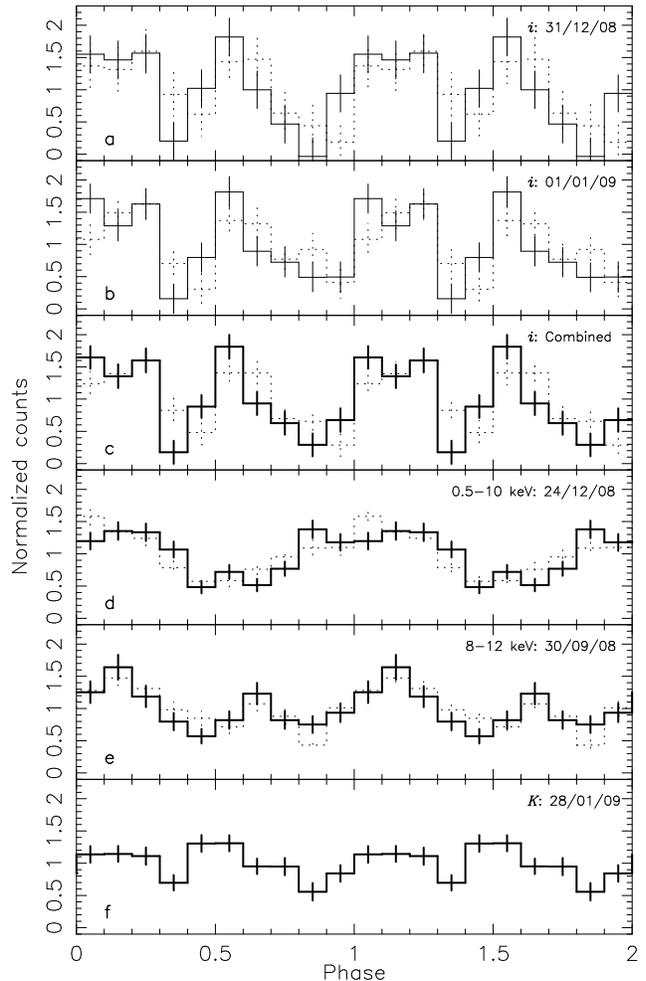}
 \caption{Pulse profiles of SGR\,0501+4516. (a) $i'$-band pulse
   profile on 2008 December 31. (b) $i'$-band pulse profile on 2009
   January 1.  (c) $i'$-band pulse profile of both nights
   combined. The solid lines and dashed lines show the pulse profiles
   obtained using techniques (i) and (ii), respectively (see
   Sections~\ref{sec:tech1}~and~\ref{sec:tech2}).  (d) $0.5-10$\,keV
   pulse profile on 2008 December 24 (solid line) and 2009 January 7
   (dashed line). (e) $8-12$\,keV pulse profile on 2008 September 30
   (solid line) and 2008 August 29 (dashed line). (f) $K$-band pulse
   profile on 2009 January 28. With the exception of the $K$-band
   light curve, which has had a phase offset of 0.1 added (see
   Section~\ref{sec:lc_ir} for details), all of the above profiles have been
   folded on the same X-ray ephemeris and hence their phases relative
   to each other are correct. For clarity, two cycles of each pulse
   profile are shown, normalised by dividing by the mean number of
   counts.}
\label{fig:lc}
\end{figure}

The first data reduction technique (Section~\ref{sec:tech1}) resulted
in the light curves shown as solid lines in
Figures~\ref{fig:lc}a,~\ref{fig:lc}b~and~\ref{fig:lc}c. The light
curves clearly show a double-peaked pulse profile, with a broad peak
around phase 0.15 and a narrower peak around phase 0.55 (where phase 0
is arbitrarily defined as MJD\,54701.0 in the X-ray ephemeris of
\citealt{rea09b}). The fact that the profile exhibits approximately
the same morphology, phasing and pulsed fraction on both nights is
strong evidence that the optical pulsations are real.

\begin{figure}
 \centering
 \includegraphics[clip,width=7.5cm,angle=270]{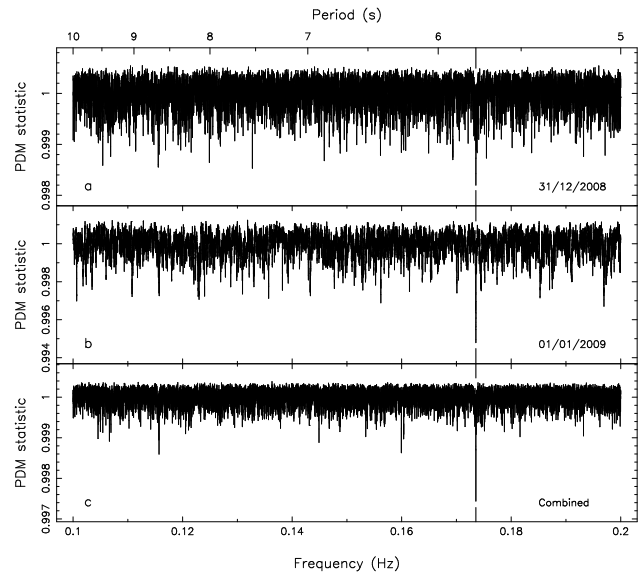}
 \caption{Phase Dispersion Minimization (PDM) periodograms of the
   $i'$-band light curve of SGR\,0501+4516. From top to bottom, the
   PDM periodogram on 2008 December 31, 2009 January 1 and both nights
   combined. The extended tickmarks at a frequency of 0.1735\,Hz show the period of the X-ray
   pulsation.}
  \label{fig:pdm}
\end{figure}

An independent test of the reality of the optical pulsations can be
obtained from the second data reduction technique
(Section~\ref{sec:tech2}). Rather than folding the data on the X-ray
ephemeris, we can instead determine the pulse period directly from our
optical data using a periodogram. The non-sinusoidal nature of the
pulsation profile led us to use the Phase Dispersion Minimization
(PDM) technique \citep{stellingwerf78} to construct a periodogram,
where the light curve is folded on a set of trial frequencies into 10
phase bins and the normalised variance of the data in each binned
light curve is then plotted as a function of frequency. The results
are shown in Figure~\ref{fig:pdm}c. The deepest trough, and hence most
likely period, in the periodogram occurs at $5.7622\pm0.0003$\,s
(where the error is given by the width ($\sigma$) of a Gaussian fit to
the trough), in excellent agreement with the X-ray period of
$5.7620692\pm0.0000002$\,s \citep{rea09b}. The corresponding light
curves are shown as dashed lines in
Figures~\ref{fig:lc}a,~\ref{fig:lc}b~and~\ref{fig:lc}c and are in good
agreement with the solid lines obtained using the first data reduction
technique, lending additional confidence to our reduction and analysis
techniques.

To test the robustness of the optical period, we constructed
periodograms of each night separately
(Figures~\ref{fig:pdm}a~and~\ref{fig:pdm}b); the deepest trough on
each night is indeed coincident with the X-ray period. We also
constructed 1000 randomised light curves from the original
16\,988-point light curve by randomly re-ordering the time-series:
none of the resulting 1000 PDM periodograms showed a lower trough than
the one at 5.7622\,s. To check for instrumental artefacts, we
calculated the periodograms of the sky, the comparison star, object
position and object FWHM -- none showed any evidence for a periodicity
at 5.7622\,s. As a final test, we also calculated a sine-fitting based
(Lomb-Scargle) periodogram of the light curve \citep{press89} -- as
expected the strongest peak occurs at precisely half the period
determined from the PDM periodogram. No other significant signals at
shorter or longer periods are present in the periodograms, and there
is no evidence for optical bursts in the light curve at levels over
4$\sigma$ above the mean.

\subsection{X-ray light curves}
\label{sec:lc_xray}

The evidence outlined above is conclusive -- we have detected optical
pulsations from SGR\,0501+4516 on the same period as the X-ray
pulsations. To compare the optical and X-ray pulse profiles, we show
$0.5-10$\,keV data obtained with the {\em Swift} X-ray telescope on
2008 December 24 and 2009 January 7 in Figure~\ref{fig:lc}d and
$8-12$\,keV data obtained with the {\em XMM-Newton} X-ray telescope on
2008 August 29 and 2008 September 30 in Figure~\ref{fig:lc}e (see
\cite{rea09b} and references therein). Like the optical light curves,
the X-ray data have been referred to the Solar System barycentre and
folded on the X-ray ephemeris of \cite{rea09b}, which means that the
relative optical/X-ray phasing is correct. The $0.5-10$\,keV X-ray
pulse profile has a broad, single-humped morphology centred around
phase 0.1, in stark contrast to the double-peaked optical profile. The
broad optical peak occurs at approximately the same phase as the
$0.5-10$\,keV X-ray peak, with the secondary optical maximum more or
less coincident with the X-ray minimum. The harder X-rays between
$8-12$\,keV, on other hand, look remarkably similar to the optical
light curves.

\subsection{Infrared light curve}
\label{sec:lc_ir}

As part of a study of the long-term infrared light curve (Levan et
al., in preparation), we also obtained some high-speed $K$-band
photometry of SGR\,0501+4516. These data were obtained with the Near
Infrared Imager (NIRI) on the 8.1\,m Gemini-North telescope in Hawaii
on 2009 January 28, when the SGR had a magnitude of $K = 19.7\pm 0.1$
(c.f. $K = 19.2\pm0.2$ on 2008 August 25; \citealt{rea08}). The
individual frames were reduced in the standard fashion using the NIRI
pipeline, utilizing a sky frame to remove quadrant offsets, and
subtracting a column-stacked median in order to remove low-level
striping in the data. Object counts were then extracted in the manner
outlined in Section~\ref{sec:tech2}. The resulting light curve,
consisting of 1168 data points each of 1\,s exposure time and 6\,s
dead time, was folded into 10 phase bins using the same ephemeris
that was used to fold the optical and X-ray data
\citep{rea09b}. However, unlike ULTRACAM, {\em Swift-BAT} and {\em
  XMM-Newton}, the accuracy of the absolute time-stamping of
Gemini-NIRI data frames is not known, and hence the absolute phase of
the $K$-band light curve is uncertain. We therefore added a phase
offset of 0.1 (i.e. $\sim$0.6\,s) to the light curve in order to align
the two minima around phases 0.35 and 0.85 with the corresponding
minima in the optical light curves. The result is shown in
Figure~\ref{fig:lc}f.  Assuming this phase offset is justified, it can
be seen that the infrared light curve shows pulsations with a similar
morphology to the optical light curves, albeit with a lower pulsed
fraction: The rms pulsed fractions (see \cite{dhillon05} for the
definition used) of the light curves presented in Figure~\ref{fig:lc}
are 52$\pm$7\%, 30$\pm$4\%, 26$\pm$5\% and 20$\pm$5\% in the optical,
0.5$-$10\,keV X-rays, 8$-$12\,keV X-rays and infrared, respectively.

We believe that our detection of infrared pulsations is robust -- a
periodogram analysis similar to the one we performed with the optical
data (Section~\ref{sec:lc_opt}) shows a peak at exactly half the X-ray
rotation period with a strength that lies in the top 0.7\% of all the
peaks in the periodogram. Although this is not significant enough to
claim an independent detection of the period in the infrared (there
are still many stronger peaks in the periodogram), it does lend
support to the reality of the observed infrared pulsations.

\section{Discussion and conclusions}
\label{sec:disconc}

Our detection of optical pulsations on the X-ray spin period has
confirmed that the proposed optical and infrared counterpart reported
by \cite{fatkhullin08} and \cite{tanvir08} is indeed SGR\,0501+4516.
This makes it the first SGR with an
unambiguously-detected\footnote{SGR\,1806$-$20 (\citealt{kosugi05},
  \citealt{israel05}) and SGR\,1900+14 \citep{testa08} have faint,
  unconfirmed infrared counterparts in crowded fields, based on the
  observation of long-term variability.}  counterpart at these wavelengths,
providing us with a unique opportunity to study the emission
mechanism.

The optical properties of SGR\,0501+4516 are remarkably similar to
those exhibited by the two AXPs with detected optical pulsations,
4U\,0142+61 (\citealt{kern02b}, \citealt{dhillon05}) and
1E\,1048.1$-$5937 \citep{dhillon09}. All three objects exhibit optical
light curves with broadly similar pulsed fractions and pulse profiles
that are in phase with the X-rays. This supports the widely-accepted
link between the AXPs and SGRs, and indicates that they share the same
optical emission mechanism. The pulse profiles of SGR\,0501+4516 are
also remarkably similar to Geminga, a rotation-powered, middle-aged,
radio-quiet pulsar with a spin period of 0.237\,s and a magnetic field
strength approximately two orders of magnitude lower than that
proposed for SGR\,0501+4516 \citep{bignami96}. Like SGR\,0501+4516,
Geminga exhibits double-peaked optical and hard X-ray pulse profiles
and a single-peaked modulation in the soft X-rays \citep{shearer98}.

It is difficult to reconcile the results presented in
Section~\ref{sec:results} with the simplest form of the fall-back disc
model of AXPs and SGRs, as any optical emission is predicted to be due
to reprocessed X-rays from the disc \citep{perna00b} and hence should
have the same or lower pulsed fraction and a similar morphology. It is
possible to appeal to beaming effects or special geometries, so that
the X-ray pulsations we see are different to those seen by the disc,
but such models seem unlikely now that optical pulsations with similar
properties have been discovered in one SGR and two AXPs. Further
modelling would be required, however, to confirm/refute whether or not
our observations are consistent with more elaborate forms of the
fall-back disc model, such as the disc-star dynamo model
(\citealt{cheng91}; \citealt{ertan04}).

Pulsed optical emission in SGR\,0501+4516 can be explained within the
context of the magnetar model by non-thermal emission from the inner
magnetosphere (see \cite{zane11} and references therein). In this
model, the softest X-rays in the 0.5-12 keV range are thermal photons
from the neutron star surface, and the hardest X-rays in this range
are produced by resonant cyclotron scattering of the surface photons
by mildly relativistic particles moving along the closed field lines
in the inner magnetosphere. The optical and infrared emission is then
produced slightly further out in the magnetosphere by curvature
radiation emitted by relativistic electrons moving along the closed
field lines. The shared magnetospheric origin of the optical, infrared
and harder X-rays suggest that they might also share similar pulse
profiles, as opposed to the softer X-rays from the neutron star
surface. This is exactly as observed in Figure~\ref{fig:lc} and
provides strong motivation to extend the model of \cite{zane11} to
predict pulse profiles for detailed comparison with these data.

\section*{Acknowledgements}
The authors acknowledge the support of the STFC. SPL also acknowledges
the support of an RCUK Fellowship. The William Herschel Telescope is
operated on the island of La Palma by the Isaac Newton Group in the
Spanish Observatorio del Roque de los Muchachos of the Instituto de
Astrofisica de Canarias.

\bibliographystyle{mn2e}
\bibliography{abbrev,refs}

\end{document}